\documentclass[
 amsmath,amssymb,
 aps,twocolumn,
 showpacs
]{revtex4-1}

\usepackage{subfig}
\usepackage{graphicx}
\usepackage{dcolumn}
\usepackage{bm}
\usepackage{ulem}

\newcommand{\sj}[6]{
\left\{
  \begin{array}{ccc}
    #1 & #2 & #3 \\
    #4 & #5 & #6 \\
  \end{array}
\right\}
}

\newcommand{\vect}[1]{\bm{#1}}

\begin{document}

\title{Nonlinear Magneto-Optical Rotation in Rubidium Vapor Excited with Blue Light}

\author{S. Pustelny}
\email{pustelny@uj.edu.pl}
\affiliation{Institute of Physics, Jagiellonian University, {\L}ojasiewicza 11, 30-348, Krak\'ow, Poland}%
\affiliation{Department of Physics, University of California, Berkeley, CA 94720-7300, USA}

\author{L. Busaite}
\affiliation{Laser Centre, University of Latvia, 19 Rainis Boulevard, Riga, Latvia, LV-1586}%

\author{A. Akulshin}%
\affiliation{Swinburne University of Technology, PO Box 218, Hawthorn Victoria, Australia 3122}%

\author{M. Auzinsh}
\affiliation{Laser Centre, University of Latvia, 19 Rainis Boulevard, Riga, Latvia, LV-1586}%

\author{N. Leefer}
\affiliation{Helmholtz-Institut Mainz, 55128 Mainz, Germany}

\author{D. Budker}
\affiliation{Helmholtz-Institut Mainz, Johannes Gutenberg Universit\"at Mainz, 55128 Mainz, Germany}
\affiliation{Department of Physics, University of California, Berkeley, CA 94720-7300, USA}
\date{\today}

\begin{abstract}
We present experimental and numerical studies of nonlinear magneto-optical rotation (NMOR) in rubidium vapor excited with resonant light tuned to the $5^2\!S_{1/2}\rightarrow 6^2\!P_{1/2}$ absorption line (421~nm). Contrary to the experiments performed to date on the strong $D_1$ or $D_2$ lines, in this case, the spontaneous decay of the excited state $6^2\!P_{1/2}$ may occur via multiple intermediate states, affecting the dynamics, magnitude and other characteristics of NMOR. Comparing the experimental results with the results of modelling based on Auzinsh \textit{et al.}, \textit{Phys. Rev. A} \textbf{80}, 1 (2009), we demonstrate that despite the complexity of the structure, NMOR can be adequately described with a model, where only a single excited-state relaxation rate is used.
\end{abstract}

\maketitle

\section{Introduction}

Nonlinear magneto-optical rotation (NMOR) is the light-intensity dependent rotation of polarization of linearly polarized light during its propagation through a medium subject to an external magnetic field. Over the years, the effect has been extensively studied, both experimentally and theoretically \cite{Budker2002}. The research has been driven by the desire for a comprehensive understanding of the physical processes responsible for the rotation of light polarization, as well as by fundamental and practical applications of the effect. For example, a detailed understanding of the generation, evolution, and detection of quantum states in atoms, manifested at the macroscale as NMOR, led to the development of techniques enabling the manipulation for these states (quantum-state engineering) \cite{Pustelny2006,Pustelny2011Tailoring} and methods of their nondestructive measurement (quantum nondemolition measurements) \cite{Takahashi1999Quantum,Gajdacz2013Nondestructive}. The effect has also been used in investigations of the relaxation of ground-state coherences in atomic vapor \cite{Budker2005Investigation,Pustelny2006Influence,Chalupczak2013Radio}, resulting in refinement of the techniques enabling the generation of long-lived ($\gtrsim$60~s) ground-state coherences \cite{footnote1}. On a more practical side, NMOR has found applications in atomic clocks \cite{Knappe2005}, optical magnetometers \cite{Budker2007, budker2013optical}, narrow-band optical filters \cite{Cere:09}, and laser-frequency locking systems \cite{Lee2011}. An interesting area of application of NMOR is fundamental research. For example, the effect is used in the search for non-magnetic spin couplings \cite{Hunter1991, Bouchiat1995, Herczeg1999,Commins2001,Ravi2015Permanent}, and is proposed for experiments focusing on the detection of constituents of dark matter/energy \cite{Pustelny2013}.

\begin{figure}
\begin{center}
\includegraphics[width=0.9\columnwidth]{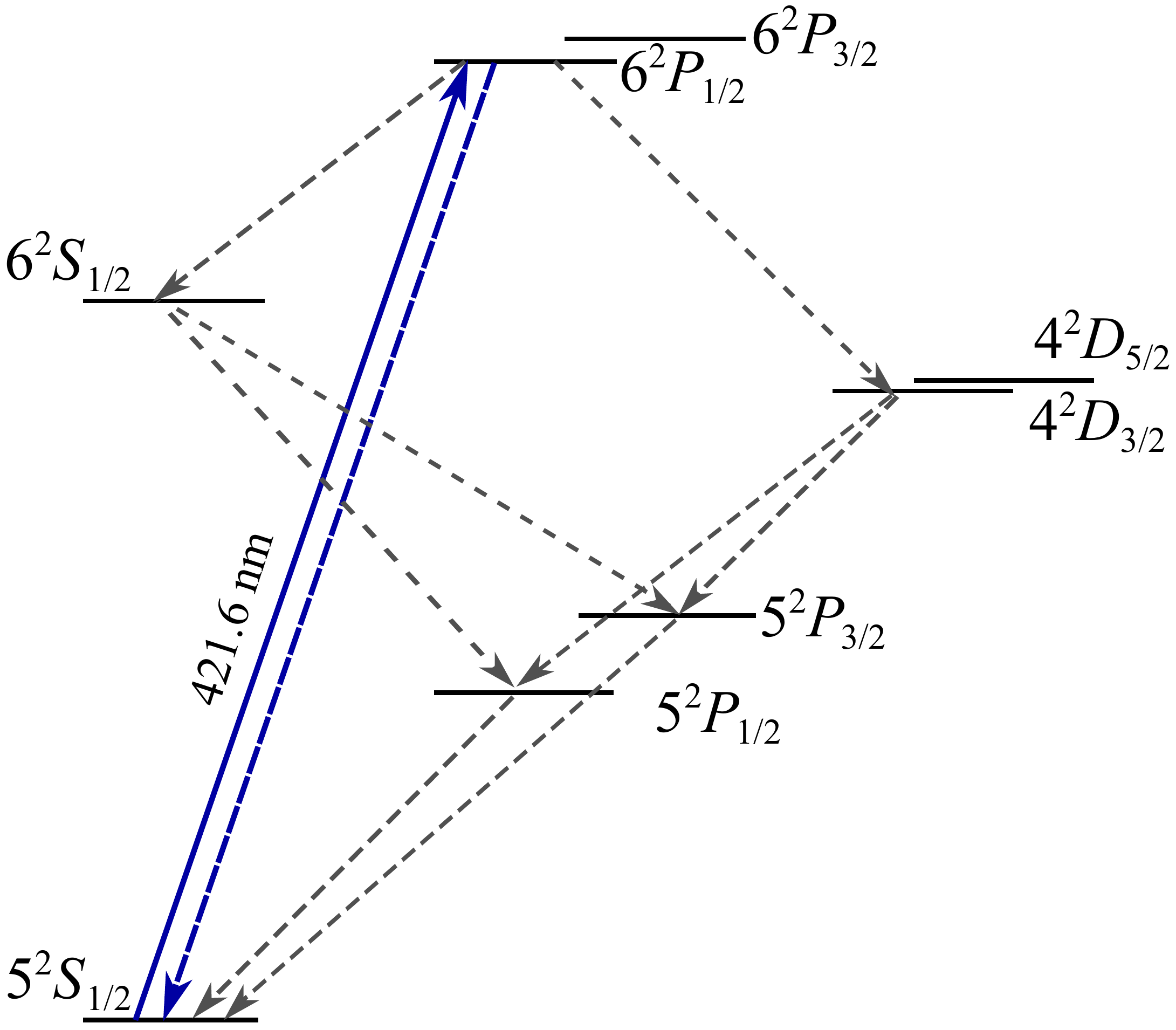}
\caption{Relevant energy states and transitions of Rb atoms excited by resonant radiation at 421.6 nm. The solid line indicate the excitation light, while the dashed lines indicate channels of spontaneous emission.}
\label{fig:levels}
\end{center}
\end{figure}

To date, most of the NMOR studies and applications utilized alkali vapors optically excited on the strong $D_1$ or $D_2$ absorption lines (see, for example, Ref.~\cite{budker2013optical} and references therein). In such systems, good agreement between theoretical predictions and experimental observations has been demonstrated \cite{Pustelny2011Tailoring,Zigdon2010Nonlinear}. The wide range of successful applications of the method resulted in a demand for its extension to different optical transitions. In this paper, we study NMOR under such different physical conditions, i.e., we explore the effect with rubidium atoms excited to a higher-energy state (the $6^2\!P_{1/2}$ state). This results in more complex repopulation of the ground-state levels; in addition to the direct repopulation of the ground-state sublevels, the repopulation my occur via several intermediate states ($6^2\!S_{1/2}$, $4^2\!D_{3/2}$, $5^2\!P_{1/2}$, and $5^2\!P_{3/2}$) (Fig.~\ref{fig:levels}) \footnote{It should be noted that nonlinear magneto-optical fluorescence resonances were recently studied theoretically and experimentally~\cite{Auzinsh2011Cascade}. In that case, cesium atoms were excited with light tuned to the $6^2\!S_{1/2}\rightarrow 7^2\!P_{3/2}$ line and fluorescent light was measured/simulated as a function of the magnetic field. This investigation revealed a good agreement between experimental observations and the results of calculations based on the optical Bloch equations.}. This enables the analysis of the role of these different relaxation channels in NMOR, while preserving other important characteristics, such as ground-state relaxation time and atomic density. Moreover, the smaller splitting of the hyperfine levels of the $6^2\!P_{1/2}$ states, relative to the $5^2\!P_{1/2}$ and $5^2\!P_{3/2}$ states, offers the possibility to investigate the role of state splitting on the efficiency of generation and probing of ground-state coherences \cite{Auzinsh2009,Auzinsh2009a}.

A specific goal of this paper is a comparison of the experimental results of the, so-called, blue NMOR, where excitation and probing at the $5^2\!S_{1/2}\rightarrow 6^2\!P_{1/2}$ absorption line is performed using 421~nm light, with the results of theoretical calculations based on the model developed in Refs.~\cite{Auzinsh2009a,Auzinsh2009}. Our aim is to verify if the real (complex but closed) system can be adequately described with a single relaxation parameter responsible for ground-state repopulation as was assumed in the model.

In addition to the understanding of NMOR in a more complex system, observation of the effect at the $5^2\!S_{1/2}\rightarrow 6^2\!P_{1/2}$ transition offers several interesting features. For example, due to the low strength of the $5^2\!S_{1/2}\rightarrow 6^2\!P_{1/2}$ transition, the excitation only weakly perturbs the atomic medium. Thus, blue NMOR may be applied as a nondestructive probe of laser cooled and trapped atoms, including quantum degenerate gases. While it is possible to perform similar measurements on stronger transitions (e.g., the $D_1$ or $D_2$ lines) using weak or detuned light, application of blue light facilitates separation of the light from the cooling/trapping (near infrared) beams, which increases precision of the state detection (optical readout is not affected by scattered light). In another application, low absorption of blue light in water provides an opportunity to use this transition in remote underwater magnetometery.

This paper is organized as follows. In the next section, we outline the theoretical approach developed in Ref.~\cite{Auzinsh2009}, recalling the most important results of the analysis (see Appendix~\ref{sec:Derivation}). In Sec.~\ref{sec:Apparatus}, we describe the experimental apparatus used for detection of blue NMOR. The experimental and theoretical results are presented and discussed in Sec.~\ref{sec:Results} and the conclusions are given in Sec.~\ref{sec:Conclusions}. Finally, the Appendix presents the derivation of the scaling parameters that enables precise calculations of blue NMOR signals and its direct comparison with experimental data.

\section{Theoretical background\label{sec:Theory}}

Nonlinear magneto-optical rotation can be considered as a three-stage process, consisting of pumping, evolution, and probing of quantum states of atoms, constituting a medium interacting with light. In the model, light first pumps atoms, modifying their optical properties and introducing medium's anisotropy. The axis of generated anisotropy is defined by the light polarization. During the next stage, the light-modified properties evolve. This evolution is caused by external fields, resulting in precession of the anisotropy around the magnetic field. The precession is accompanied by processes intrinsic to the system, which leads to relaxation of the atomic polarization and hence medium's optical anisotropy. In the final stage, modified properties of the medium affect propagation of light, leading to the rotation of light polarization. Although in real experiments these three stages typically occurs simultaneously, here, without loss of generality, we consider them independently (in Appendix~\ref{sec:Derivation} we eventually shows a polarization-rotation formula that combines all the stages).

To describe the quantum evolution of a system, one may utilize the density-matrix formalism. This approach allows determination of a quantum state of atoms interacting with light subject to an external magnetic field, i.e., the situation that we encounter in NMOR. While an explicit form of the Liouville equation governing quantum evolution of the system is provided in Appendix~\ref{sec:Derivation}, here, we only recall the most important elements of the derivation of the equations for polarization rotation.

In our approach, the light-atom interaction is treated semi-classically and is described by the electric-dipole Hamiltonian. Optical excitation is accompanied by spontaneous emission, covered in the Liouville equation by a spontaneous-emission operator. Both processes are responsible for optical pumping of the medium and hence generation of its polarization. It should be stressed, however, that despite the significant number of the intermediate states present in the system, in our model we only use a single excited-state relaxation rate. This reduces the problem to the simpler ``two''-level system. The question if this assumption is correct or is oversimplification is one of the main problems addressed with this work.

To conveniently describe optical pumping, one may use the irreducible tensor basis. This approach allows us to introduce several simplifications into the theoretical description. In particular, it can be shown that in low light-intensity regime, absorption of a photon by an unpolarized atom may only induce the three lowest polarization moments: population, orientation, and alignment (see discussion in Appendix~\ref{sec:Derivation}). Moreover, if light polarization direction is aligned along the quantization axis the only non-zero atomic polarization generated during pumping stage is the alignment population distribution ($\rho^{(2\ \!\!0)}$), while the only non-zero moment existing initially in the system, $\rho^{(0\ \!\!0)}$ moment (isotropic polarization), is modified \footnote{The $\rho^{(\kappa\ \!\!q)}$ denotes the polarization moment of the rank $\kappa$ and component $q$.}.

During the next (evolution) stage, the only external interaction present in the system arises from the magnetic field, i.e., the interaction Hamiltonian is exclusively given by the Zeeman interaction, and the relaxation. The interaction, in its lowest order, does not change the rank of the polarization moment, but it can mix components of the same-rank moments. Thereby, the polarization moments $\rho^{(2\ \!\pm\!1)}$ can be generated in the system during the evolution stage.

Rotation of light polarization occurring during the last, probing, stage can be calculated using the macroscopic polarization of the medium. This enables calculation of the medium's electric susceptibility and hence refractive indices for two orthogonal circular polarizations of light responsible for light-polarization rotation. In the considered case, the only polarization moments contributing to the effect are the $\rho^{(2\ \!\!1)}$ and $\rho^{(2\ \!-\!1)}$ polarization moments (Appendix~\ref{sec:Derivation}). Since these moments are generated during the first two stages, this shows the role of light (generation of the $\rho^{(2\ \!\!0)}$ polarization moment) and magnetic field in NMOR.

It should be stressed that our model was developed under the assumptions of low light intensities and small magnetic fields. The low light intensity ensures the absence of such nonlinear optical processes as alignment-to-orientation conversion \cite{Budker2000Nonlinear}, which would result in deterioration of NMOR-signal amplitude. The assumption of low magnetic field ensures that the effect arises due to the ground-state coherences, i.e., is not not caused by linear magneto-optical rotation or the so-called Bennet-structure effect \cite{Budker2002a}.

A specific question that needs to be addressed in the theoretical considerations of NMOR is motion of atoms. As shown in Refs.~\cite{Auzinsh2009,Auzinsh2009a}, the atomic motion should be treated differently when atoms are contained in a buffer-gas-free uncoated vapor cell, a buffer-gas-filled cell, or a paraffin-coated cell. The simplest situation is in the first case, i.e., the scenario investigated in this work, when the alkali-atom collisions are rare, and the most important mechanism of ground-state relaxation are collisions with the walls. The absence of velocity-class mixing ensures during the interrogation period ensures that detuning does not between pumping and probing, i.e., light generates and probes atomic polarization at the same hyperfine transition. Thereby, the signal from each velocity group can be calculated independently and then summed over the velocity distribution with weights given by the distribution. This allows calculation of polarization rotation from the whole ensemble.

A formula for ensemble-averaged polarization rotation [Eq.~(\ref{eq:RotationDoppler})] was used to calculate all theoretical signals compared to the experimental data throughout this paper.

\section{Experimental apparatus\label{sec:Apparatus}}

\begin{figure}[hbt!]
    \includegraphics[width=\columnwidth]{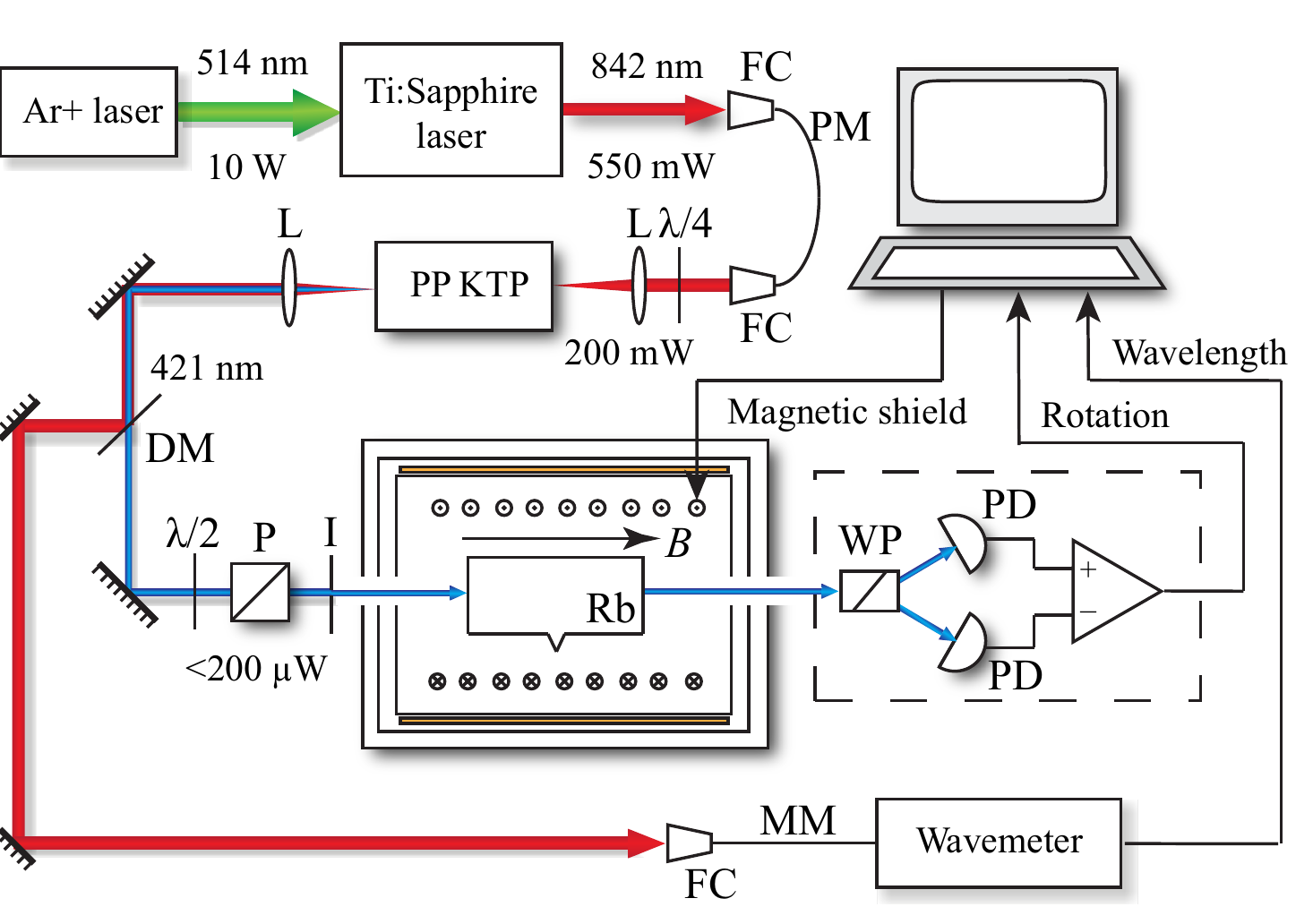}
    \caption{Experimental apparatus. P is the polarizer, WP denotes the Wollastone prism, $\lambda$/2 is the halfwave plate, L is the lens, DM denotes the dichroic mirror, I is the iris, FC stands for the fiber coupler, PD is the photodiode, and PM and MM stand for polarization-maintaining and multimode fiber, respectively.}
    \label{fig:setup}
\end{figure}

\begin{figure}[htb!]
    \begin{center}
    \includegraphics[width=\columnwidth]{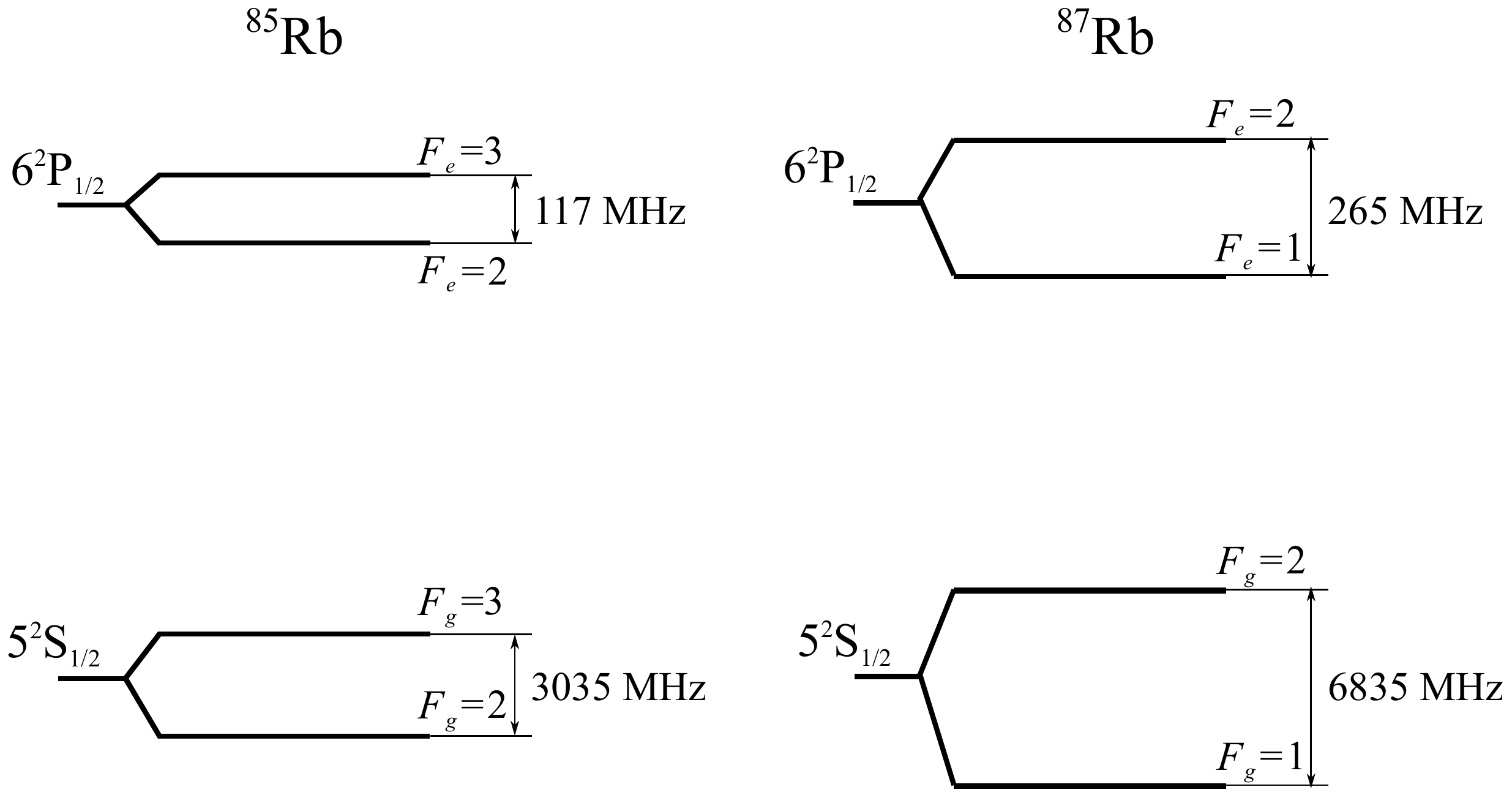}
        \caption{Energy-level diagrams of the $5^2S_{1/2}\rightarrow 6^2P_{1/2}$ absorption line in $^{87}$Rb and $^{85}$Rb.}
    \label{fig:level85}
    \end{center}
\end{figure}

A schematic of the experimental apparatus is shown in Fig.~\ref{fig:setup}.
An Ar$^+$ laser (Coherent Innova-400), emitting 10~W of continuous-wave (CW) green (514~nm) light, was used to pump a Ti:Sapphire laser (Coherent 899 ring laser). The laser enabled generation of 550~mW of near infrared (IR) light (842 nm), which was then coupled into a single-mode optical fiber and delivered to a different part of the building, where the main experiment was performed. The light coming out of the fiber was focused on a periodically poled potassium titanyl phosphate (PP KTP) crystal. Nonlinear properties of the crystal enabled conversion of the IR light (200~mW) into 0.2~mW of blue (421~nm) light in a single-pass configuration \footnote{It is worth noting that blue-light resonant with the $5^2\!S_{1/2}\rightarrow 6^2\!P_{3/2}$ absorption line could be efficiently generated in step-wise excitation of a rubidium vapor \cite{Zibrov2002}. This collimated radiation could have relatively high intensity (above 1~mW) \cite{Sell2014} and sub-MHz linewidth \cite{Akulshin2014}.}. The blue-light frequency was controlled by tuning the Ti:Sapphire laser. This enabled tuning the blue light in resonance with all transitions of the $5^2\!S_{1/2}\rightarrow 6^2\!P_{1/2}$ line of both rubidium isotopes ($^{85}$Rb and $^{87}$Rb) (see Fig.~\ref{fig:level85}). For the measurements of NMOR signal versus the light intensity, the blue light frequency was stabilized by referencing the Ti:Sapphire laser to an optical cavity and was monitored with a wavemeter (HighFinesse WS/7). An  absorption-spectroscopy setup, employing an independent reference cell containing rubidium vapor heated to 70$^\circ$C, was used as an additional reference for the measurements.

Rubidium (natural abundance) vapor, being a magneto-optically active medium, was contained in an evacuated cylindrical (50~mm in diameter and 100~mm in length) glass cell placed inside a four-layer cylindrical magnetic shield made of mu-metal. The shield offered a passive attenuation of external magnetic fields at a level better than 10$^6$ \cite{budker2013opticalYashchuk}. A set of additional coils mounted inside the shield's innermost layer was used to compensate residual magnetic fields and to apply a bias magnetic field along the light-propagation direction. The bias field was generated with a 12-bit current-output card, driving longitudinal magnetic-field coils, enabling generation of up to 1-G field. The shield layers were thermally isolated and the innermost layer was resistively heated,  along with the vapor cell within it, to about 90$^\circ$C, corresponding to a vapor density of $2.4\times10^{12}$~atoms/cm$^3$ and a Doppler width of the transition of about 600~MHz.

In front of the shield, the blue light was spectrally filtered (the IR light was directed to the wavemeter by a dichroic mirror) and it was linearly polarized with a high-quality crystal polarizer. The laser-beam diameter was controlled by a iris \footnote{Note that opening the iris not only changes the beam diameter but also its profile.}. After the cell, the polarization of the light was detected with a balanced polarimeter, consisting of a Wollaston polarizer and two photodiodes. The photodiodes' differential photocurrent divided by twice the sum of the photocurrents provides a rotation angle $\varphi$ of NMOR [$\varphi=(I_1-I_2)/(2I_1+2I_2)$ for $\varphi\ll 1$]. The signal was stored with a computer, which was also used to control the card generating the current for the magnetic coils inside the magnetic shield.

\section{Results and Discussion\label{sec:Results}}

\begin{figure}[tb!]
    \includegraphics[width=\columnwidth, trim = 0mm 30mm 0mm 30mm]{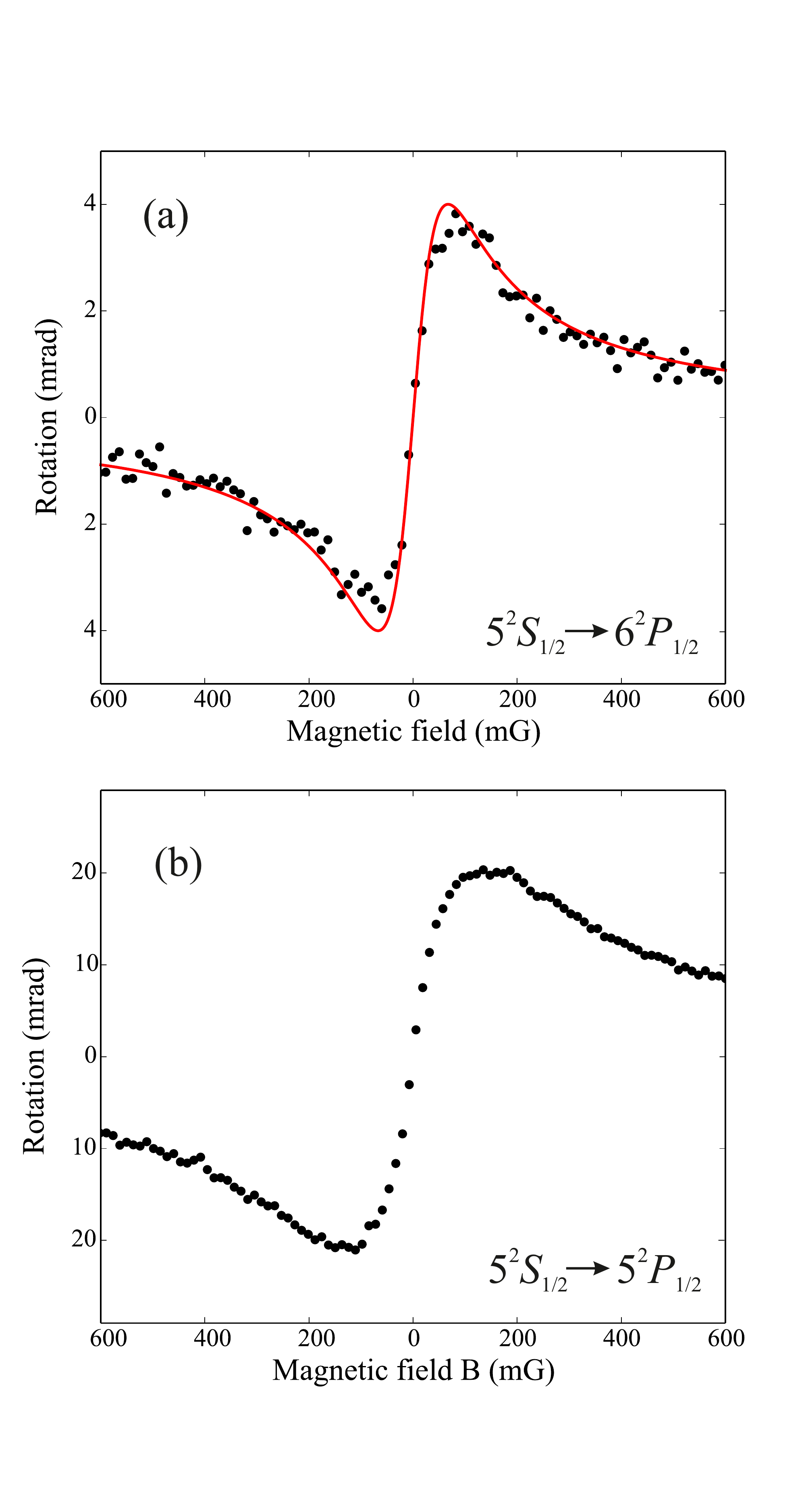}
    \caption{Rotation of linearly polarized blue (a) and IR (b) light tuned to the $^{85}$Rb $F_g=3\rightarrow F_e$ transition of the $5^2\!S_{1/2}\rightarrow 6^2\!P_{1/2}$ line and the $^{85}$Rb $F_g=3\rightarrow F_e$ transition of the $5^2\!S_{1/2}\rightarrow 5^2\!P_{1/2}$ line, respectively. In both measurements, light intensities and beam diameters were the same (1.9~mW/cm$^2$ and 1~mm, respectively), while temperature of the cell was 85$^\circ$C for blue-light measurements and 23$^\circ$C for IR measurements. The solid red curve superimposed on the data shown in (a) is the theoretical curve calculated using parameters extracted from the experiment (see text).}
    \label{fig:NMORSignal}
\end{figure}

Figure~\ref{fig:NMORSignal}(a) shows a typical NMOR signal measured with blue light.
Similarly, as in NMOR observed at the $D_1$ line [Fig.~\ref{fig:NMORSignal}(b)], the signal is centered at zero magnetic field and reveals a dispersive shape. For a given set of parameters, the blue-NMOR signal has amplitude of 4~mrad and width of 80~mG (width is defined as the magnetic-field difference between maximum and minimum of the signal.) While this rotation was measured at roughly $85^\circ$C, 5 times larger rotation was observed on the $D_1$ line at 23$^\circ$C, corresponding to nearly 3 orders of magnitude lower concentration. In both cases, the lasers were tuned to the same ground state (the $F=2$ state) and they had identical intensities. This clearly shows the difference in strength of magnetic-field-induced optical anisotropy for these two absorption lines \footnote{At low absorption depths, NMOR scales linearly with atomic density number.}.

\begin{figure}[ht!]
    \includegraphics[width=\columnwidth, trim = 10mm 0mm 10mm 10mm]{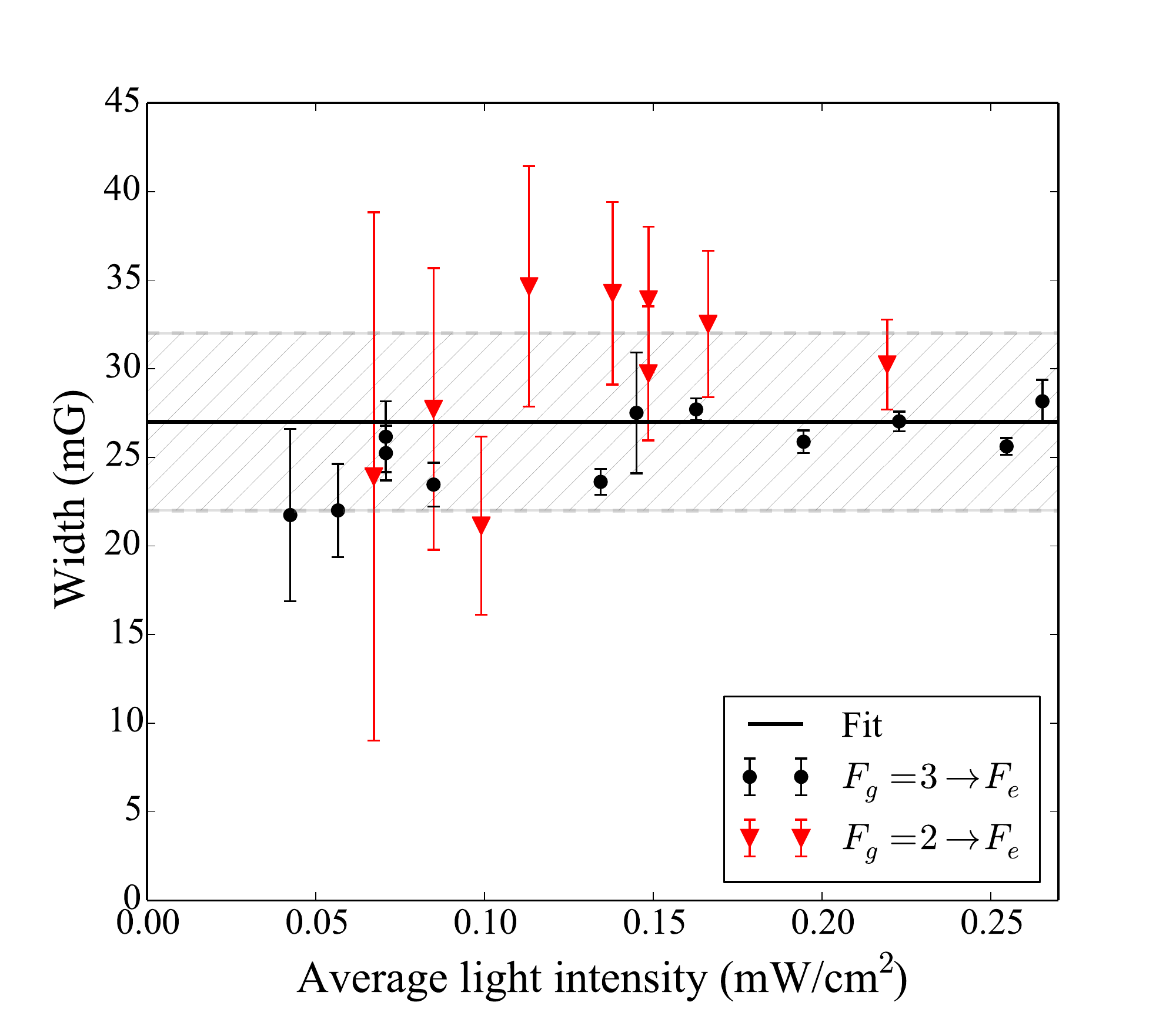}
    \caption{Widths of blue-NMOR resonances measured as a function of average light intensity for two transitions ($F_g=3\rightarrow F_e$ - black circles and $F_g=2\rightarrow F_e$ - red triangle) of the $5^2\!S_{1/2}\rightarrow 6^2\!P_{1/2}$ line. The error bars of the $F_g=2\rightarrow F_e$ points are much larger than those in the $F_g=3\rightarrow F_e$ case, as the former signals are much smaller, i.e., with worse signal-to-noise ratio. The solid black line is the constant value fitted to the experimental data with the uncertainties marked with the shaded area. Absence of the power broadening of the both resonances proves the weakly perturbing character of the interaction. The signals were measured at 85$^\circ$C with beam diameter of 6~mm. The shaded area marks the uncertainty in the determination of the parameter $\kappa_\gamma$.}
    \label{fig:WidthVsPower}
\end{figure}

The capabilities of blue NMOR as a weakly perturbing probe most prominently manifest themselves when the widths of resonances observed at the two transitions are compared; the NMOR resonance observed at the $D_1$ line is nearly two times broader than the one observed at the $5^2\!S_{1/2}\rightarrow 6^2\!P_{1/2}$ transition. This difference originates from power broadening, which at the $D_1$ line dominates the resonance width. The absence of power broadening is confirmed with the measurements of the width of the blue-NMOR resonance versus area-average light intensity (Fig.~\ref{fig:WidthVsPower}) \footnote{While in the experiment the intensity reveals a specific spatial profile, we present NMOR signals as a function of the area-average light intensity. The real beam profile is replaced by a top-hat intensity profile of the same diameter and total light power.}.
As shown, the width of the signal remains constant in the entire accessible range of light intensities. These results confirm the applicability of our theoretical model developed in a low-light-intensity regime \footnote{From the intensity range used in the current work and known dipole matrix elements of the $5^2S_{1/2}\rightarrow 6^2P_{1/2}$ line, we determined a Rabi-frequency range of 0--1.2~MHz.}.

Next, we compare the experimental data with the results of simulations. A correct description of the problem requires determination of several experimental parameters: the Larmor frequency $\Omega_{F_g}$, Doppler broadening $\Gamma_D$,  Rabi frequency $\Omega_R$, and ground-state relaxation rate $\gamma$. While determination of the first two quantities is relatively straightforward based on the temperature of the vapor, strength of the magnetic field, and the excitation scheme, the estimation of the remaining two parameters is more challenging.

In the considered case of an evacuated buffer-gas-free vapor cell and absence of the power broadening, the ground-state relaxation rate is exclusively determined by the time of flight of atoms across the light beam. In the naive approach, one would expect that the rate is equal to the mean transverse (to the light-propagation direction) velocity of atoms over the mean path across the light beam \footnote{In the case of a cylindrically-shaped beam, the mean path length across the light beam of the diameter $d$ is equal to the beam radius $r$.}. However, as shown in Ref.~\cite{Pflenghaar1993}, the effect of transit of atoms of different velocity groups across the light beam is more complex. For instance, the atoms that spend more time within the light beam (atoms with smaller transverse velocities) contribute more pronouncedly to the NMOR signal, causing essential narrowing of the signal \cite{Pflenghaar1993}. To accommodate for this effect, we modified the naive formula by introducing the parameter $k_\gamma$ so that the relaxation rate takes the form
\begin{equation}
\gamma = k_{\gamma}\frac{v_{tr}}{r},\label{eq:gamma}
\end{equation}
where $r$ is the beam radius and $v_{tr}=\sqrt{k_BT/m}$ is the average transverse velocity, with $T$ being the temperature, $m$ denoting the mass of the atom, and $k_B$ being the Boltzmann constant. The value of $k_\gamma=0.42(7)$ for a 6-mm beam was determined based on the fitting of the Lorentz curve to the experimental data (Fig.~\ref{fig:WidthVsPower}). This value was then kept constant for all light intensities and light tunings \footnote{It is noteworthy that $k_\gamma$ varies with beam diameters. This is understandable based on the discussion of Ref.~\cite{Pflenghaar1993}; the change of the beam diameter modifies the weight of the contribution of the atoms from different velocity groups and hence changes the overall width of the NMOR signal.}. The parameter for other beam diameters is given in Table~\ref{table:Diameters}.
\begin{table}[b!]
    \caption{Parameters $k_\gamma$ and $k_\Omega$ calculated and determined based on the experimental data for different apertures of the iris situated in front of the cell.}
    \begin{center}
        \begin{tabular}{|c|c|c|c|}
          \hline
          $d$~(mm) & $k_\gamma$ & $k_\Omega^{Cal}$ & $k_\Omega^{Exp}$ \\ \hline
          1 & 0.17(1) & 1.02 & 1.04(1) \\
          3 & 0.23(2) & 1.15 & 1.30(2) \\
          6 & 0.42(7) & 1.60 & 1.62(7) \\
          \hline
        \end{tabular}
    \end{center}
    \label{table:Diameters}
\end{table}

Another parameter that needs to be determined for the calculation is the Rabi frequency. While the Rabi frequency can be determined from the light intensity, the correct modeling of NMOR requires incorporation of the effective Rabi frequency. This originates from the fact that the light beam does not have a top-hat intensity profile and atoms traversing the beam experience different values of the Rabi frequency across the light beam. Therefore, in order to correctly describe the system, we introduce the parameter $k_\Omega$, accommodating for the effect. This parameter effectively replaces a real beam intensity profile with a top-hat profile, providing the Rabi frequency in a form
\begin{equation}
    \Omega_R = k_\Omega\frac{d E_0}{\hbar},
    \label{eq:OmegaR}
\end{equation}
where $d$ is the electric dipole matrix element and $E_0$ is the amplitude of the electric field of light, which can be calculated from the light intensity $I$ [$I=cE_0^2/(8\pi)$]. The parameter $k_\Omega$ can be calculated by integrating the Rabi frequency over the actual beam profile (see Appendix~\ref{sec:AppendixA} for more details) \cite{Auzinsh2015}.

Table \ref{table:Diameters} presents the parameter $k_\Omega^{Cal}$ calculated numerically based on the beam profile and iris apperture and the parameter $k_\Omega^{Exp}$ arising from the fitting of the Lorentz curve to the experimental data. As shown, the values of the parameters depend on opening of the iris (and hence beam diameter), as the aperture determines the profile of the beam. In particular, with a fully open aperture (6 mm), atoms interact with a beam of the Gaussian profile, while closing the aperture to 1~mm changes the profile to nearly top-hat shape. Thereby, in the first case, the actual beam profile strongly deviates from the top-hat shape and the value of $k_\Omega$ is significantly larger than one, while in the second case it is nearly one.

Determining $k_\gamma$ and $k_\Omega$ enables one to simulate the blue-NMOR signals. Figure~\ref{fig:NMORSignal}(a) presents the experimental data overlaid with results of theoretical calculations. The simulated curve reproduces the features of the experimental data including the signals' shapes, amplitudes, and widths. It should be stressed that besides the parameter $k_\gamma$, the agreement between experimental results and theoretical simulations was achieved with parameters directly extracted from the experiment.

\begin{figure}[tb!]
    \includegraphics[width=\columnwidth, trim = 0mm 0mm 10mm 10mm]{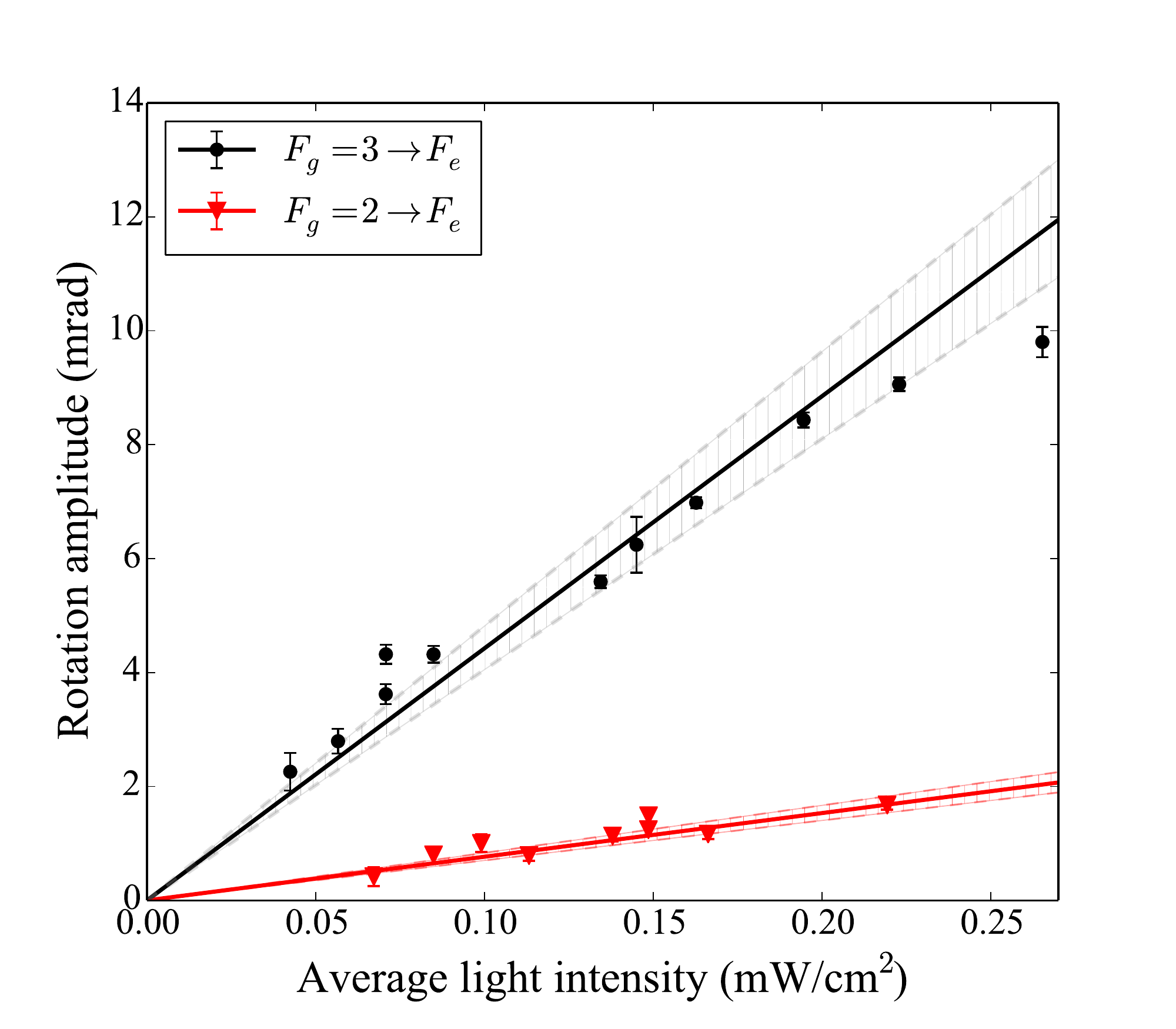}
    \caption{Amplitudes of NMOR signals measured versus the average intensity of light tuned to different ground-state hyperfine states of $^{85}$Rb. The experimental data are in agreement with theory. The signals were measured at 85$^\circ$C with 6-mm beam diameter. The shaded areas mark the uncertainty due to the limited precision of $\kappa_\Omega$-parameter determination.}
    \label{fig:AmplitudeVsPower}
\end{figure}

To further check the theoretical model, we studied the dependence of the NMOR signals on the light intensity. As shown in Fig.~\ref{fig:WidthVsPower}, in the accessible intensity range the width of blue-NMOR signal is independent on the intensity. At the same time, the nonlinear character of NMOR implies the dependence of the amplitude of the signal on the intensity. Figure~\ref{fig:AmplitudeVsPower} shows the amplitude of blue NMOR signal as a function of the averaged light intensity measured at two transitions of $^{85}$Rb ($F_g=3\rightarrow F_e$ and $F_g=2\rightarrow F_e$) (Fig.~\ref{fig:level85}).
The presented data sets reveal different dependences on the average light intensity. For a given beam diameter (i.e., fixed $k_\gamma$ and $k_\Omega$), this difference is well reproduced by theory (the difference stems from different dipole matrix elements associated with the transitions).

\begin{figure}[htb!]
    \includegraphics[width=\columnwidth, trim = 0mm 30mm 10mm 30mm]{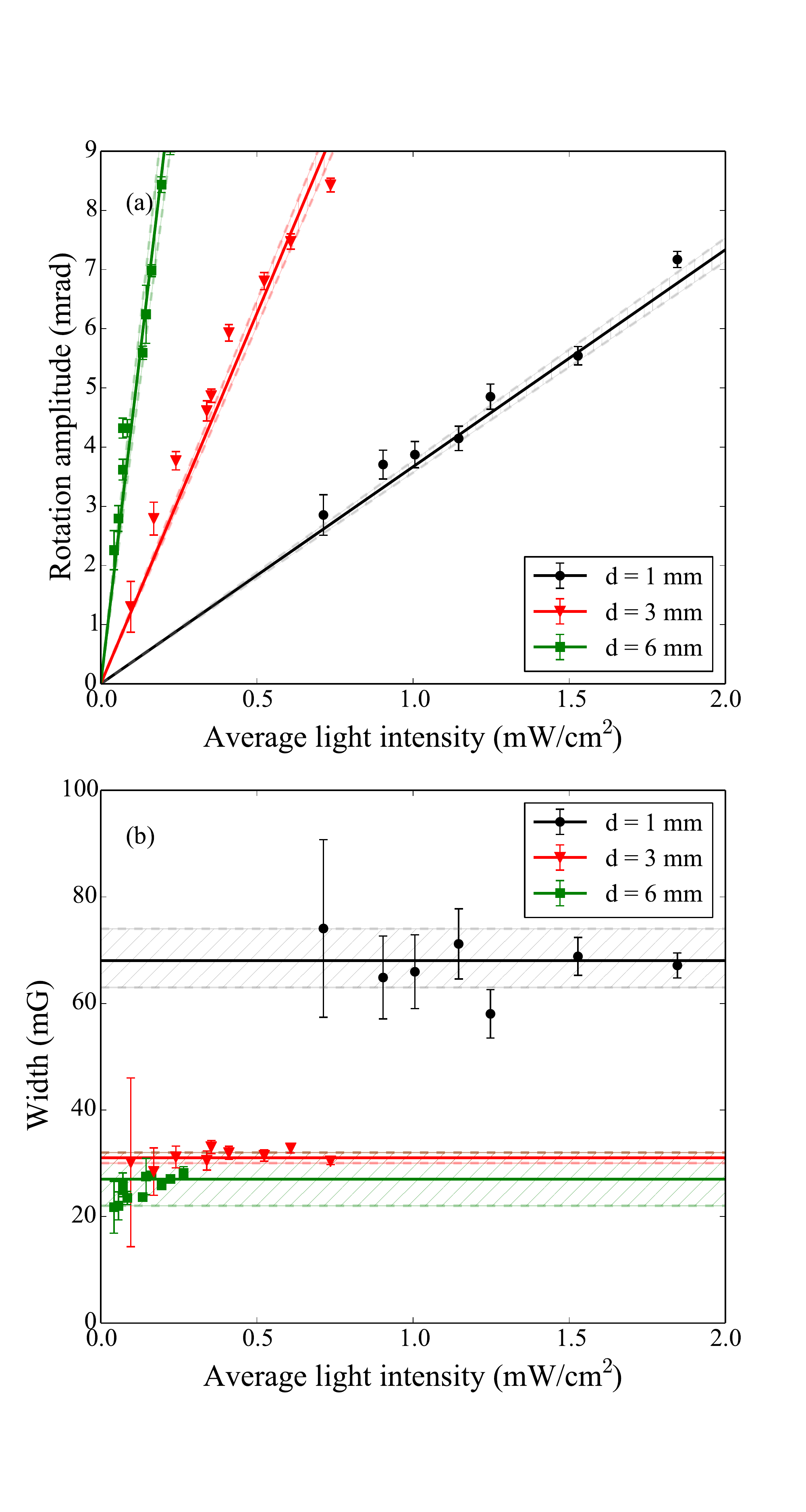}
    \caption{Amplitude (a) and width (b) of blue NMOR signal measured versus average light intensity for three iris apertures (i.e., for different beam diameters and profiles). All three experimental data sets reveal good agreement with the theory. The signals were measured for the laser tuned to the $F_g=3\rightarrow F_e$ transition of $^{85}$Rb.}
    \label{fig:NMORVsDiameter}
\end{figure}

The dependence of the NMOR signal on beam diameter is shown in Fig.~\ref{fig:NMORVsDiameter}.
Analysis of the amplitude and width versus the average light intensity reveals some interesting features of the dependences. First is the difference in the slopes of the amplitude dependence on the average light intensity for three iris apertures [3.7(1)~{mrad$\times$cm$^2$/mW}, 12.5(4)~{mrad$\times$cm$^2$/mW}, 22(2)~{mrad$\times$cm$^2$/mW} for 1-mm, 3-mm, and 6-mm openings, respectively]. This originates from the different ground-state relaxation rate in the three cases [see Fig.~\ref{fig:NMORVsDiameter}(b)] and hence  different values of the saturation parameter $\kappa_s$. However, to adequately reproduce the dependences, one also needs to take into account the different beam profiles of the beams [getting into the saturation parameter via the parameters $k_\Omega$ (Table~\ref{table:Diameters})]. Second, the width of the resonance, although independent of the light intensity, depends of the iris opening [Fig.~\ref{fig:NMORVsDiameter}(b)]. Interestingly, the widths are not in straightforward relation with the diameter of the iris opening. The experimentally determined widths for 3-mm and 6-mm openings are the same within the error bars [31(1)~mG and 27(4)~mG, respectively], the width for the smallest opening is roughly 2 times broader [68(5)~mG]. As the blue-NMOR signals are not power broadened, this difference is not related with the effect and it stems from the beam profile \footnote{As the beam diameter is 4 mm, opening the iris from 3~mm to 6~mm does not affect the reak light-intensity profile strongly.} but also is a consequence of the different dynamics of the optical pumping for distinct beam diameters \cite{Pflenghaar1993}.

The final step in our studies of blue NMOR was the investigation of the NMOR spectra.
\begin{figure}[hbt!]
    \includegraphics[width=\columnwidth, trim = 0mm 0mm 0mm 0mm]{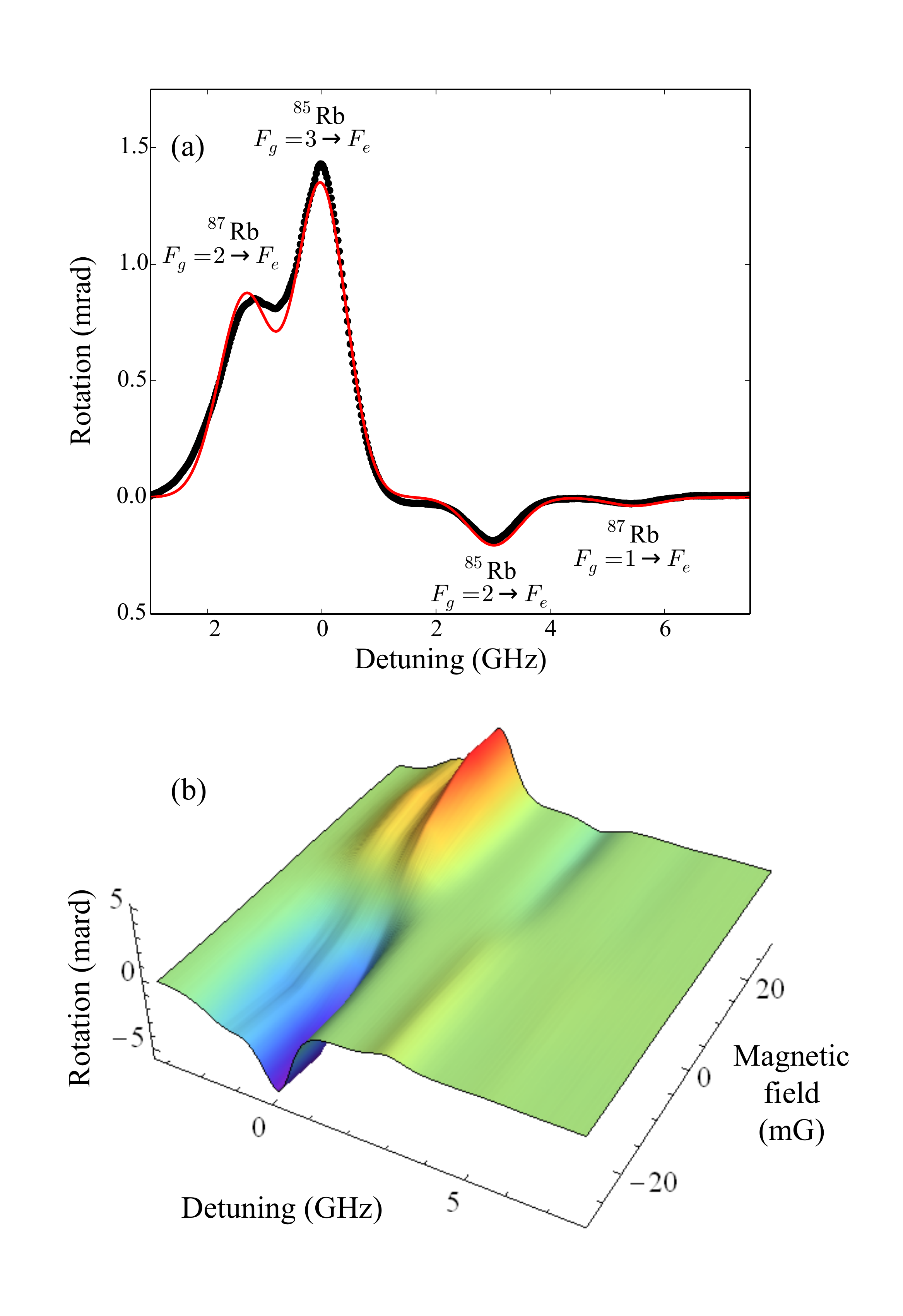}
    \caption{(a) Measured and simulated NMOR spectra recorded for a magnetic field of about 8~mG and light intensity of 0.1~mW/cm$^2$ and (b) rotation spectra measured for different magnetic fields and light intensity of 0.1~mW/cm$^2$. The signals were obtained for light frequency scanned across all transitions of the $5^2\!S_{1/2}\rightarrow 6^2\!P_{1/2}$ line.}
    \label{fig:NMORspectra}
\end{figure}
Figure~\ref{fig:NMORspectra} shows rotation of polarization as a function of blue-light detuning for the whole $5^2\!S_{1/2}\rightarrow 6^2\!P_{1/2}$ line of $^{85}$Rb and $^{87}$Rb. The signal was measured at a magnetic field of roughly 20~mG and light intensity of ${\sim\!0.1}$ mW/cm$^2$. As discussed in Sec.~\ref{sec:Theory}, the absence of buffer gas and paraffin-coating of the walls results in the absence of ``communication'' between atoms from different velocity classes. Therefore, atoms from each of the velocity classes contributes independently to the signal and the theoretical signal may be simulated by simply summing the contributions from atoms of different velocity groups. We apply this approach for calculating NMOR spectra. An example of such simulations is presented in Fig.~\ref{fig:NMORspectra}(a) along with the experimental data. As shown, the agreement of the simulations with the experimental data is good. For example, the calculation and measurements show that for a given magnetic field the strongest rotation was observed at the $F_g=3\rightarrow F_e$ transition of $^{85}$Rb \footnote{It should be stressed that an additional parameter contributing to the ratio of rotation observed at different rubidium isotopes is the Land\'e factor; even in the power-broadening-free regime (same ground-state relaxation rate $\gamma$), i.e., maximum rotation is observed at different magnetic fields. Therefore, the ratio between the NMOR signal observed at the $^{85}$Rb and $^{87}$Rb transition may vary depending on the magnetic field.}. This rotation was roughly 2 times larger than the amplitude of the second largest peak observed at the $F_g=2\rightarrow F_e$ transition of $^{87}$Rb. The rotation ratio at these two transitions is roughly the same as the ratio of absorption at the transitions. Much weaker signals were observed at the other transitions (the $F_g=2\rightarrow F_e$ transition of $^{85}$Rb, and the $F_g=1\rightarrow F_e$ transition of $^{87}$Rb.) In contrast to the former case, however, the difference in NMOR-signal amplitudes is much larger than the difference in the absorption at the lines. Moreover, the signs of the rotation at these two lines is reversed with respect to the rotation observed at the $F_g=3\rightarrow F_e$ and $F_g=2\rightarrow F_e$ transitions of $^{85}$Rb and $^{87}$Rb, respectively. The opposite sign stems from the opposite Land\'e factor of the two ground states. These observations were confirmed with the spectra measured at different magnetic field for slightly smaller light intensity (0.1~mW/cm$^2$ with respect to 0.2~mW/cm$^2$) [Fig.~\ref{fig:NMORspectra}(b)].

\section{Conclusions\label{sec:Conclusions}}

We presented results on nonlinear magneto-optical rotation at the $5^2\!S_{1/2}\rightarrow 6^2\!P_{1/2}$ line (421~nm) of both stable isotopes of rubidium. In contrast to NMOR measured on the $D_1$ or $D_2$ line, excited atoms in the considered case may decay through several intermediate states. Generally, this modifies the dynamics of repopulation pumping, potentially changing the characteristics of NMOR signals and their dependences on various experimental parameters. With our results we demonstrated that the theoretical description agrees with experimental data, reproducing such characteristics of NMOR signals as their amplitudes, widths, and spectral dependences. This agreement was achieved with the model where only one excited-state relaxation rate was used. This is an interesting result as it was achieved in the system where only $15\%$ of atoms return directly to the ground state, while remaining 85\% reach the state via several intermediate states.

The agreement between the theory and NMOR experiment with a buffer-gas-free uncoated vapor cell implies investigation of the effect in systems with velocity group mixing, being more sensitive to hyperfine splittings of the states. This can be realized in a buffer-gas-filled or high-temperature paraffin-coated cell. The investigations would further test the model developed in Ref.~\cite{Auzinsh2009}.

Beyond the fundamental understanding of NMOR induced by light coupling ground states with higher excited states, our results demonstrate the potential of such an excitation scheme for magnetometry. This would be particularly interesting in remote underwater magnetometry, due to weak absorbtion of blue light in water \footnote{Absorption of blue light in water is roughly three orders of magnitude smaller than absorption of ~800-nm light with the strong $D_1$ or $D_2$ lines (see, for example, Ref.~\cite{Budker2000Nonlinear}).}.

\section{Acknowledgements}

The authors would like to thank Simon Rochester for his role in the development of the theoretical model of NMOR. LB and MA acknowledge support from ERAF project Nr. 2010/0242/2DP/2.1.1.1.0/10/APIA/VIAA/036, NATO Science for Peace and Security Programme project SfP983932, and the Latvian Council of Science project 119/2012. SP acknowledges support from the National Centre for Research and Development within the Leader programme.

\appendix
\section{Calculating the parameter $k_\Omega$\label{sec:AppendixA}}

\begin{figure}[b!]
    \includegraphics[width=\columnwidth]{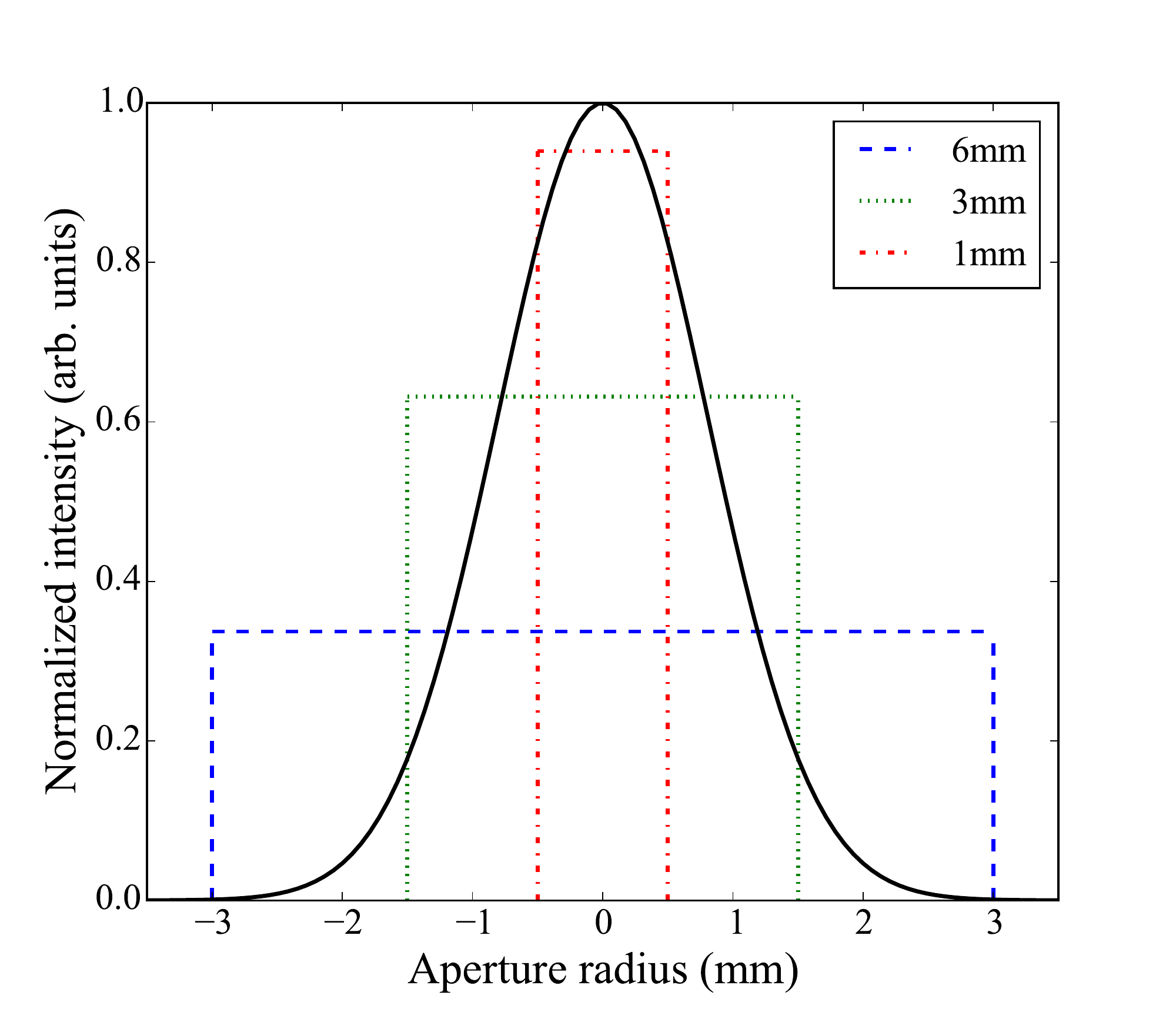}
    \caption{Schematic depicting the effect of an iris on the light-intensity profile of light used in the experiment. For the iris openings significantly larger than the beam diameter (half-width at half maximum,) the real profile strongly deviates from the top-hat function. At the same time, the small openings result in a profile close to the top-hat profile.}
    \label{fig:GaussianBeam}
\end{figure}

To estimate the parameter $k_\Omega$, we consider a light beam of a Gaussian intensity profile
\begin{equation}
I = I_0 e^{-\frac{2 r'^2}{\sigma^2}},
\end{equation}
where $I_0$ is the maximum intensity at the center of the beam and $\sigma$ is the beam radius at which intensity drops $1/e^2$ before the iris. When the beam reaches the iris, the iris cuts part of the beam off, modifying its beam diameter and intensity profile (Fig.~\ref{fig:GaussianBeam}). In such a case, the total light power is then given by
\begin{equation}
P = 2\pi I_0 \int_0^{R_a} \! r'e^{- \frac{2r'^2}{\sigma^2}} \, \mathrm{d}r',
\label{eq:RealIntensityProfile}
\end{equation}
where $R_a$ is the radius of the iris aperture. Analogically, the top-hat intensity profile used in our model is
\begin{equation}
P = I_1\pi R_a^2,
\label{eq:TopHatIntensityProfile}
\end{equation}
where $I_1$ is the light intensity. In order to substitute the real light-intensity profile [Eq.~(\ref{eq:RealIntensityProfile}] with the same radius and power top-hat intensity profile [Eq.~\ref{eq:TopHatIntensityProfile}], one needs to introduce the normalization parameter $k_\Omega$. This parameter can be determined by
\begin{equation}
k_\Omega = \sqrt{\frac{I_0}{I_{1}}} = \sqrt{\frac{\pi R_a^2}{\int_0^{R_a} \! r'e^{-\frac{2 r'^2}{\sigma^2}} \, \mathrm{d}r'}}.
\label{eq:kOmega}
\end{equation}
It should be stressed that the parameter only depends on iris diameter and is independent of the light power, thus we keep it constant for all light intensites of a given beam diameter.

The key experimental parameter that is required to calculate $k_\Omega$ is the beam radius $\sigma$. From the measurements of the beam profile, we determined the radius as 2~mm. This value enabled to calculate the parameters $k_\Omega$ for all beam diameters (see Table~\ref{table:Diameters}).

\section{Model derivation\label{sec:Derivation}}

Dynamics of a quantum system can be described using the density-matrix formalism. In the considered case of atoms contained in a vapor cell subject to external fields (e.g., light and magnetic field), the time evolution of the density matrix $\rho$ is governed by the Liouville equation ($\hbar=1$, $c=1$)
\begin{equation}
    \dot{\rho}=-i[H,\rho]-\frac{1}{2}\{\Gamma,\rho\}+\Lambda+\textrm{Tr}(S\rho).
\label{eq:Liouville}
\end{equation}
Here, $H=H_0+V$ is the Hamiltonian of the system, consisting of the unperturbed Hamiltonian $H_0$ and the interaction operator $V$, $\Gamma$ is the relaxation operator, $\Lambda$ is the repopulation operator describing density-matrix-independent repopulation processes such as those induced by wall collisions, and $S$ is the repopulation operator, which describes processes depending on the density matrix, e.g., spontaneous emission.

In the context of NMOR, it is useful to present quantum operators, including the density matrix $\rho$, in the irreducible tensor basis. In that case, a quantum operator $A$ can be written as
\begin{equation}
    A=\sum_{\kappa q} A^{\kappa q}_{\widetilde{F}\widetilde{F}'}T^{(\kappa)}_{q;\widetilde{F}\widetilde{F}'},
    \label{eq:OperatorExpanssion}
\end{equation}
where, $T_{q;\widetilde{F}\widetilde{F}'}^{(\kappa)}$ is the $q$-th component of the irreducible tensor $T^{(\kappa)}$ with rank $\kappa$ ($\kappa=0,...,2F$ and $q=-\kappa,...,\kappa$), and $A^{\kappa q}_{\widetilde{F}\widetilde{F}'}$ are the expansion coefficient called polarization moments (PMs) or multipoles. The subscripts $\widetilde{F}$ and $\widetilde{F}'$ indicate the set of quantum numbers enabling unambiguous determination of the state with the total angular momentum $F$ and $F'$ that the operator acts on. The polarization moments $A^{\kappa q}_{\widetilde{F}\widetilde{F}'}$ of the operator $A$ are related to the set of the matrix elements $A_{Fm,F'm'}$ of the operator $A$ in the $|Fm\rangle$ basis, where $m$ and $m'$ are the magnetic quantum number, through the Wigner-Eckart theorem
\begin{widetext}
    \begin{equation}
        A^{(\kappa q)}_{\widetilde{F}\widetilde{F}'}=\textrm{Tr}\!\left[A\left(T^{(\kappa)}_{q;\widetilde{F}\widetilde{F}'}\right)^+\right]=
            \sum_{mm'}(-1)^{F'-m'}\langle FmF'-\!m'|\kappa q\rangle A_{Fm,F'm'}.
        \label{eq:BasisTransformation}
    \end{equation}

Using Eq.~(\ref{eq:BasisTransformation}), one can expand Eq.~(\ref{eq:Liouville}) in the irreducible tensor basis, which for the ground state takes the form
    \begin{eqnarray}
        \dot{\rho}^{(\kappa q)}_{\widetilde{F}_g\widetilde{F}'_g}&=&i(-1)^{F_g+F_g'+\kappa+1}\sum_{\kappa',\kappa'',q',q'',F''}\sqrt{[\kappa'][\kappa'']}\langle\kappa'q'\kappa''q''|\kappa q\rangle\sj{\kappa'}{\kappa''}{\kappa}{F_g'}{F_g}{F''}\times \nonumber \\
        &\times&\left[\left(H^{(\kappa'q')}_{\widetilde{F}_g\widetilde{F}''}-\frac{i}{2}\Gamma^{(\kappa'q')}_{\widetilde{F}_g\widetilde{F}''}\right)\rho^{(\kappa''q'')}_{\widetilde{F}''\widetilde{F}_g'}-\rho^{(\kappa'q')}_{\widetilde{F}_g\widetilde{F}''}
        \left(H^{(\kappa''q'')}_{\widetilde{F}''\widetilde{F}_g'}+\frac{i}{2}\Gamma^{(\kappa''q'')}_{\widetilde{F}''\widetilde{F}_g'}\right)\right]+\nonumber\\
        &+&\Lambda^{(\kappa q)}_{\widetilde{F}_g\widetilde{F}_g'}+\frac{4}{3}\omega_0^3\sum_{F_e,F_e'>F_g,F_g'}\langle \widetilde{F}_g\|d\|\widetilde{F}_e\rangle \rho^{(\kappa q)}_{\widetilde{F}_e\widetilde{F}'_e}\langle \widetilde{F}_e'\|d\|\widetilde{F}_g'\rangle (-1)^{F_e+F_e'+\kappa+1}\sj{\kappa}{F'_g}{F_g}{1}{F_e}{F_e'},
        \label{eq:rhoKq}
    \end{eqnarray}
\end{widetext}
where $\langle \widetilde{F}_g\|d\|\widetilde{F}_e\rangle$ is the reduced dipole matrix element given by
\begin{widetext}
    \begin{equation}
        \langle \widetilde{F}_g\|d\|\widetilde{F}_e\rangle=(-1)^{J_e+I+F_g+1}\sqrt{[F_g][F_e]}\sj{J_e}{F_e}{I}{F_g}{J_g}{1}\langle J_g\|d\|J_e\rangle
        \label{eq:ReducedDipoleMatrix}
    \end{equation}
\end{widetext}
with $I$ being the nuclear spin and  $J$ being the electron angular momentum. For the compactness of the formula above, the symbol $[x]=2x+1$ was introduced. $\langle J_g\| d\| J_e\rangle$ appearing in Eq.~(\ref{eq:ReducedDipoleMatrix}) can be calculated from
\begin{equation}
    R\gamma_e=\frac{4}{3}\frac{\omega_0^3}{2J_e+1}\langle J_g\|d\|J_e\rangle^2,
\end{equation}
where $R$ is the branching ratio for the transition $J_e\rightarrow J_g$, $\gamma_e$ is the excited-state relaxation rate, and $\omega_0$ is the transition frequency between ground and excited states. The first two terms of Eq.~(\ref{eq:Liouville}) are accommodated in Eq.~(\ref{eq:rhoKq}) by the expression under the sum. The second term in Eq.~(\ref{eq:rhoKq}) arises due to density-matrix-independent repopulation, while the last term comes from the repopulation due to the spontaneous emission.

To induce the polarization rotation of light, an atom first needs to be polarized/pumped with light, next the polarization has to evolve due to the external magnetic field and finally this magnetic-field-modified polarization has to affect the polarization of light. While in real systems these stages occur simultaneously, to calculate polarization rotation, it is useful to consider them independently.

Equation~(\ref{eq:rhoKq}) can be used to calculate the evolution of the density matrix during the first pumping stage, when atomic polarization is generated, but also during the second stage when the polarization evolves due to the magnetic field. The difference between these two stages manifests in the form of the interaction operator $V$, which, in the first case, is determined by the electric-dipole interaction between light and atoms and in the second case by the magnetic (Zeeman) interaction. In turn, the interaction operator at the two stages takes the form
\begin{eqnarray}
    V^{pump}&=&-\vect{E}\cdot\vect{d},\\
    V^{evol}&=&-\vect{\mu}\cdot\vect{B},
\end{eqnarray}
where $\vect{E}$ is the electric field of light, $\vect{B}$ is the magnetic field, $\vect{d}$ is the electric dipole moment, and $\vect{\mu}$ is the magnetic dipole moment. In the description of the evolution stage, we also drop the last term of Eq.~(\ref{eq:rhoKq}) as there is no optical pumping.

To calculate rotation of the linear polarization of light traversing the medium, one may use the wave equation
\begin{equation}
    \frac{\partial^2\vect{E}}{\partial l^2}-\frac{\partial^2\vect{E}}{\partial t^2}=4\pi\frac{\partial^2 \vect{P}}{\partial t^2},
\end{equation}
where $l$ is the distance along the light propagation direction and the relation of the medium's dipole polarization $\vect{P}$ with the density matrix $\rho$ is given by $\vect{P}=N_\textrm{at}\textrm{Tr}\left(\rho\vect{d}\right)$, where $N_\textrm{at}$ is the atomic density. The polarization $\vect{P}$ can be related to the density matrix $\widetilde{\rho}$ written in the rotating frame via
\begin{equation}
    \vect{P}=N_\textrm{at}\sum 2\textrm{Re}(\widetilde{\rho}_{F_gm_g,F_em_e}\bm{d}_{F_em_e,F_gm_g}) e^{i(\vect{k}\cdot\vect{r}-\omega t)},
\end{equation}
where $\vect{d}_{F_gm_g,F_em_e}$ is the electric-dipole matrix element between the ground state $|F_gm_e\rangle$ and excited state $|F_em_e\rangle$ (the sum runs over all ground- and excited states), $\widetilde{\rho}_{F_gm_g,F_em_e}$ is the corresponding rotating-frame density-matrix element (optical-coherence amplitude), and $\omega$ is the light frequency. Based on this approach, it can be shown that polarization rotation of light is given by
\begin{widetext}
    \begin{equation}
        \frac{d\varphi}{d\ell}=-\frac{4\pi\omega N_{\textrm{at}}}{\varepsilon_0}\sum_{F_gF_em_gm_e}\textrm{Im}\left[\rho'_{F_gm_g,F_em_e}\bm{d}_{F_em_e,F_gm_g}\cdot(\hat{\bm{k}}\times\hat{\bm{e}})\right],
        \label{eq:rotationGeneral}
    \end{equation}
\end{widetext}
where $\hat{\bm{k}}$ is the unit vector of the wave vector $\bm{k}$ and $\hat{\bm{e}}$ is the light-polarization unit vector.

To make use of the formulas above, the electric-dipole moment $\bm{d}_{F_em_e,F_gm_g}$ and the density-matrix element $\widetilde{\rho}_{F_gm_g,F_em_e}$ need to be presented in the irreducible tensor basis. Then, Eq.~(\ref{eq:rotationGeneral}) takes the form \cite{Auzinsh2009}
\begin{equation}
    \frac{d\varphi}{d\ell}=-2\pi\omega N_{\textrm{at}}\textrm{Im}\left[\hat{\bm{e}}\cdot\overleftrightarrow{\bm{\beta}}\cdot(\hat{\bm{k}}\times\hat{\bm{e}})\right],
    \label{eq:rotationkq}
\end{equation}
where the tensor $\overleftrightarrow{\bm{\beta}}$ is given by
\begin{widetext}
    \begin{equation}
        \overleftrightarrow{\bm{\beta}}=\sum_{F_gF_e\kappa q'q''}\frac{(-1)^{F_g+F_e+\kappa}}{\widetilde{\omega}_{F_eF_g}}\hat{\bm{e}}_{-q'}\hat{\bm{e}}_{-q''} \langle 1q'1q''|\kappa\ q'\!\!+\!q''\rangle\sj{1}{1}{\kappa}{F_g}{F_g}{F_e}|\langle F_g\|d\|F_e\rangle|^2\rho^{(\kappa\ q'\!+q'')}_{\widetilde{F}_g\widetilde{F}_g},
        \label{eq:beta}
    \end{equation}
\end{widetext}
with $\hat{\bm{e}}_q$ being the spherical-basis unit vectors. It should be noted that  Eq.~(\ref{eq:rotationkq}) was derived within the first-order perturbation theory for the optical coherence. In the derivation, the ground-state coherences between different hyperfine states were neglected.

To explicitly relate the polarization rotation [Eq.~(\ref{eq:rotationkq})] with the system parameters but to present it in a simpler analytic form several assumptions, arising from the fact that a photon is a spin-1 particle, may be introduced. Thereby, absorption of a photon by an unpolarized atom may only result in generation of the three lowest PMs ($\rho^{(\kappa)}_{q}$ with rank $\kappa\leq 2$) called population ($\kappa=0$), orientation ($\kappa=1$), and alignment ($\kappa=2$). Moreover, in the case of interaction of unpolarized atoms with light linearly polarized along the quantization axis, the symmetry of the problem indicates that only the zero component of the atomic polarization ($q=0$) may be generated. Finally, since there is no preferred direction but only a preferred axis, as a result of light-atom interaction only the polarization moment $\rho^{(20)}$ can be generated in the system (the explicit formula for $\rho^{(20)}$ is given in Refs.~\cite{Auzinsh2009a,Auzinsh2009}), while simultaneously the polarization moment $\rho^{(00)}$ is being modified in the interaction.

Using the same symmetry arguments as above, it can be shown that rotation of light polarization can be only caused by the PMs with rank $\kappa=1$ or $\kappa=2$. In fact, analysis of Eq.~(\ref{eq:beta}) reveals that the only PM affecting the polarization rotation is $\rho^{(2\ \!\pm\!1)}_{\widetilde{F}_g\widetilde{F}_g}$. Taking this into account, one can calculate the rotation of the polarization plane of $\bm{z}$-polarized light traversing an atomic sample with completely resolved hyperfine structure ($\omega_{F_gF_g'}\gg \gamma_e$,
$\omega_{F_eF_e'}\gg \gamma_e$) \footnote{The solution for the case, where this condition is not fulfilled is shown in Ref.~\cite{Auzinsh2009,Auzinsh2009a}.}
\begin{widetext}
    \begin{eqnarray}
        \ell_0\frac{d\varphi}{d\ell}&=&\frac{3}{4}\kappa_s\sum_{F_gF_e}(-1)^{2F_g}\frac{[F_e]^3[F_g]^3}{[I]}\sj{1}{1}{2}{F_g}{F_g}{F_e} {\sj{J_e}{F_e}{I}{F_g}{J_g}{1}}^4x_{F_g}[L(\omega'_{F_eF_g})]^2\times\nonumber\\
        &\times&\left(\frac{(-1)^{2I+2J_g}}{[F_e][F_g]}\sj{1}{1}{2}{F_g}{F_g}{F_e}+R(-1)^{F_g-F_e}(2J_e+1)\sj{1}{1}{2}{F_e}{F_e}{F_g}\sj{F_g}{F_g}{2}{F_e}{F_e}{1}{\sj{J_e}{F_e}{I}{F_g}{J_g}{1}}^2\right)\!\!,
        \label{eq:RotationSingleAtom}
    \end{eqnarray}
\end{widetext}
where $\ell_0=\left(1/\mathcal{I}\times d\mathcal{I}/d\ell\right)^{-1}$ is the unsaturated resonant absorption length. The Lorentz profile $L(\omega'_{F_eF_g})$ is given by
\begin{equation}
    L(\omega'_{F_eF_g})=\frac{(\gamma_e/2)^2}{(\gamma_e/2)^2+\omega_{F_eF_g}'^2},
\end{equation}
with $\omega'_{F_eF_g}$ being the Doppler shifted transition frequency of a pair of hyperfine states ($\omega'_{F_eF_g}=-\Delta\omega_{F_eF_g}+\bm{k}\cdot\bm{v}$, where $\Delta\omega_{F_eF_g}$ is the detuning of light from the optical transition, $\bm{v}$ denotes the velocity of atoms, interacting with the light), $x_{F_g}$ is the magnetic-lineshape parameter given by
$$
    x_{F_g}=\frac{(\gamma_g/2)\Omega_{F_g}}{(\gamma_g/2)^2+\Omega_{F_g}^2}.
$$
Here, $\Omega_{F_g}=g_{F_g}\mu_BB$ is the Larmor frequency of the ground-state hyperfine level $F_g$ ($g_{F_g}$ is the Land\'e factor of the ground state $F_g$ and $\mu_B$ is the Bohr magneton), $\gamma_g$ is the ground-state relaxation rate, and $\kappa_s=\langle J_g\|d\|J_e\rangle^2\varepsilon^2_0/(\gamma_g\gamma_e)$ is the reduced optical saturation parameter.

Although Eq.~(\ref{eq:RotationSingleAtom}) concerns motion of atoms by introducing the Doppler shift, the equation only concerned atoms from a single velocity group. To properly describe the atomic ensemble, one needs to include the velocity distribution of atoms in the sample. In the case of atoms contained in buffer-gas-free uncoated vapor cell, polarization of an optically pumped atom is preserved until a collision of the atom with a cell wall \footnote{Atomic concentration is so low that the mean free path of an atom is orders of magnitude larger than the beam diameter.}. Thus, atomic velocity and hence light detuning do not change between pumping and probing stages and light generates and probes atomic polarization generated in a given hyperfine state. In such a case, the signal from each velocity group can be calculated independently and then summed over the velocity distribution with weights given by the distribution, leading to the ensemble-averaged polarization rotation \cite{Auzinsh2009,Auzinsh2009a}
\begin{widetext}
    \begin{eqnarray}
        \left\langle\ell_0\frac{d\varphi}{d\ell}\right\rangle_v&=&\frac{3}{4}\kappa_s\sum_{F_gF_e}(-1)^{2F_g}\frac{[F_e]^3[F_g]^3}{[I]}\sj{1}{1}{2}{F_g}{F_g}{F_e} {\sj{J_e}{F_e}{I}{F_g}{J_g}{1}}^4e^{\Delta_{F_eF_g}/\Gamma_D}x_{F_g}\times\nonumber\\
        &\times&\left(\frac{(-1)^{2I+2J_g}}{[F_e][F_g]}\sj{1}{1}{2}{F_g}{F_g}{F_e}+R(-1)^{F_g-F_e}(2J_e+1)\sj{1}{1}{2}{F_e}{F_e}{F_g}\sj{F_g}{F_g}{2}{F_e}{F_e}{1}{\sj{J_e}{F_e}{I}{F_g}{J_g}{1}}^2\right)\!\!,\nonumber\\
        &&
        \label{eq:RotationDoppler}
    \end{eqnarray}
\end{widetext}
where
\begin{equation}
\ell_0 = -\left(\frac{1}{\mathcal{I}}\frac{d\mathcal{I}}{d\ell}\right)^{-1} = \frac{4\sqrt{\pi}}{RN_\textrm{at}\lambda^2}\frac{\Gamma_D}{\gamma_e}\frac{(2J_g+1)}{(2J_e+1)}, \label{abs_lenght}
\end{equation}
with $\lambda$ being the wavelength and $\Gamma_D$ the Doppler broadening. Since our experiment is performed in a buffer-gas-free uncoated vapor cell, we use Eq.~(\ref{eq:RotationDoppler}) to simulate blue NMOR signals.

\bibliography{NMOR_paper_reference}

\end{document}